\documentclass[sigconf]{acmart}

\usepackage{calc}
\usepackage{array}
\usepackage{booktabs}
\providecommand{\tightlist}{%
  \setlength{\itemsep}{0pt}\setlength{\parskip}{0pt}}
\setcopyright{none}
\begin{document}

\title{Spike-Killer: Evidence-Gated LLM Assistance for Safe Performance Diagnosis on a Real Windows Workstation}

\author{Baocheng Zeng}
\affiliation{%
  \institution{Tsinghua University}%
  \city{Beijing}%
  \country{China}%
}

\author{Jinhao Yang}
\affiliation{%
  \institution{Tsinghua University}%
  \city{Beijing}%
  \country{China}%
}

\begin{abstract}
LLM-assisted agents can synthesize system evidence, propose configuration changes, and automate diagnostic tasks, but their flexibility makes an imprecise action or an intrusive collector an operational risk. We present Spike-Killer, a human-approved workflow for diagnosing frame-time complaints on one real Windows workstation. The workflow treats each action as an evidence-gated transaction: it records the exact target state, classifies risk, preserves a snapshot, verifies a postcondition, and retains failed measurements as first-class evidence.

This experience paper reports a completed same-day study with Counter-Strike 2 as a demanding target application. The evidence bundle contains preserved state snapshots, exploratory microbenchmarks, a ten-run same-state repeatability probe, live telemetry, a repaired over-broad registry action, incompatible presentation-capture attempts, an invalid local replay, and a system-level tracing replacement. Windows Performance Recorder produced two CS2 local-Bot GPU traces of 90.69 and 85.85 seconds; both were attributed to \texttt{cs2.exe}, exposed DxgKrnl Present metadata, and had zero lost ETW buffers or events. These results qualify trace integrity, not performance: the study reports no frame intervals, P99 estimate, or intervention effect. The contribution is an auditable, human-in-the-loop pattern for trustworthy agent assistance on a real workstation, including explicit stop conditions when evidence is insufficient.
\end{abstract}



\maketitle

\hypertarget{introduction}{%
\section{Introduction}\label{introduction}}

A gaming laptop can feel uneven even while reporting a high average
frame rate. Average FPS compresses a time series into one throughput
value and can conceal a small number of long frames. Controlled
perceptual work has shown that variation in frame timing changes
perceived smoothness in first-person games \cite{klein2023}. Tail frame
time, rather than average FPS alone, is therefore the relevant outcome
for an intermittent-stutter complaint.

The cause may sit outside the game. A capture overlay can hook
presentation, login applications can create short resource bursts, a
game may render at the wrong resolution, and an OEM power plan can trade
responsiveness for energy savings. These mechanisms cross application,
user, driver, and privileged operating-system boundaries. Conventional
online ``optimization'' lists often bundle them, omit evidence and
rollback, and evaluate success with one before/after screenshot. An LLM
can help connect heterogeneous evidence to plausible mechanisms and
generate the glue code needed to inspect them. The same flexibility is
dangerous when its output mutates a real machine: a syntactically valid
command can resolve to the wrong object, and a measurement tool can
perturb or destabilize its target.

Spike-Killer asks a deliberately narrow question: can safety-constrained
LLM assistance make frame-time troubleshooting on \emph{one fixed
laptop} more auditable and experimentally testable within one day? We do
not compare models, claim autonomy, or generalize to other computers. We
use the July 2026 troubleshooting history as a same-day case and
explicitly separate three levels of evidence:

\begin{itemize}
\tightlist
\item
  \textbf{supported now:} the workflow was exercised across system
  layers; background resource state changed; exploratory system and
  live-match telemetry were collected; the operator reported symptom
  relief; and material action/measurement failures were observed and
  preserved;
\item
  \textbf{not supported now:} a reduction in CS2 P99 frame time,
  attribution of any performance change to an individual intervention,
  collector non-interference, or agent safety without supervision;
\item
  \textbf{out of scope for this paper:} a multi-day randomized crossover
  that estimates an intervention effect on CS2 P99 frame time.
\end{itemize}

This pilot/experience paper makes four contributions. First, it
describes a transaction-like, safety-constrained workflow for
LLM-assisted performance maintenance. Second, it reports a traceable
same-day single-machine case without converting subjective relief or
synthetic throughput into a frame-time claim. Third, it analyzes an
over-broad system mutation and measurement failures as first-class
outcomes. Fourth, it releases executable checks that verify a loss-free,
CS2-attributed WPR trace without exporting private game contents.

\hypertarget{background-and-related-work}{%
\section{Background and Related
Work}\label{background-and-related-work}}

\hypertarget{frame-timing-and-performance-diagnosis}{%
\subsection{Frame timing and performance
diagnosis}\label{frame-timing-and-performance-diagnosis}}

Frame rate and frame time are related but not interchangeable summaries.
Long or irregular inter-frame intervals can be perceptible even when
average throughput remains high; Klein et al.~experimentally manipulated
variable frame timing and found an effect on perceived smoothness
\cite{klein2023}. This motivates P99 and long-frame-rate endpoints in a
future effect study, rather than selecting a result after inspecting
many FPS summaries.

Frame-time diagnosis is also an observer-effect problem. Presentation
tracing, overlays, graphics API capture, hardware capture, and
application-controlled timing observe different points in the rendering
pipeline and impose different costs. A familiar interface around the
same underlying collector is not an independent fallback. Spike-Killer
therefore treats the collector as an intervention: it must pass
responsiveness, focus, completeness, and data-quality gates before it
can produce publication data.

\hypertarget{agents-operating-real-computers}{%
\subsection{Agents operating real
computers}\label{agents-operating-real-computers}}

OSWorld evaluates computer-use agents in executable desktop environments
and emphasizes reproducible initial states plus execution-based
validation \cite{osworld2024}. That perspective is useful here, but
performance engineering differs from ordinary desktop completion: the
desired state is statistical, the machine is thermally
history-dependent, and privileged actions may have delayed side effects.
A command returning success is not sufficient evidence of either a
correct state transition or a performance benefit. Spike-Killer narrows
the task to one machine and makes exact postconditions, rollback, and
repeated measurement part of success.

Recent safety benchmarks such as OS-Harm separately evaluate whether
computer-use agents can cause harm even when prompts are not overtly
malicious \cite{osharm2025}. Our experience supplies a complementary
in-situ lesson: ordinary performance-maintenance intent can still
produce an unsafe target scope. We do not infer population-level safety
rates from one incident; we use it to derive enforceable guards.

\hypertarget{scope-of-single-system-evidence}{%
\subsection{Scope of single-system
evidence}\label{scope-of-single-system-evidence}}

A same-day single-machine experience study answers a narrower question
than an N-of-1 performance trial: whether the workflow, its safety
guards, and its measurement path operate on the target computer. It
cannot establish a persistent intervention effect. N-of-1 reporting
guidance \cite{vohra2015} and randomized crossover guidance
\cite{dwan2019} remain useful references for a later extension, but no
such crossover is reported here. We retain raw failures, validation
outputs, and session-level provenance so that later repeated work can
begin from an auditable state.

\hypertarget{spike-killer-design}{%
\section{Spike-Killer Design}\label{spike-killer-design}}

\hypertarget{transactional-loop}{%
\subsection{Transactional loop}\label{transactional-loop}}

Spike-Killer organizes each optimization into eight stages:

\begin{enumerate}
\def\labelenumi{\arabic{enumi}.}
\tightlist
\item
  \textbf{Observe:} collect game configuration, process, startup,
  driver, display, capture, power, and workload evidence.
\item
  \textbf{Hypothesize:} state a mechanism that links the observed
  condition to tail latency.
\item
  \textbf{Classify:} label privilege, risk, reversibility, and
  human-confirmation requirements.
\item
  \textbf{Snapshot:} preserve the exact original value or file before
  mutation.
\item
  \textbf{Act:} apply the narrowest available change at the lowest
  sufficient privilege.
\item
  \textbf{Validate:} query the exact postcondition rather than trusting
  command exit status.
\item
  \textbf{Measure:} repeat the frozen workload and retain raw
  observations, including invalid attempts.
\item
  \textbf{Decide:} retain, restore, or mark the action inconclusive
  based on safety and evidence.
\end{enumerate}

This separates a \emph{plausible diagnosis}, a \emph{state-changing
action}, and a \emph{measured effect}. The agent may propose the first
two, but only controlled measurement supports the third.

\hypertarget{intervention-layers}{%
\subsection{Intervention layers}\label{intervention-layers}}

The artifact separates four reversible layers that build cumulatively on
a restorable baseline:

\begin{center}
\footnotesize
\begin{tabular}[]{@{}
  >{\raggedright\arraybackslash}p{(\columnwidth - 6\tabcolsep) * \real{0.25}}
  >{\raggedright\arraybackslash}p{(\columnwidth - 6\tabcolsep) * \real{0.25}}
  >{\raggedright\arraybackslash}p{(\columnwidth - 6\tabcolsep) * \real{0.25}}
  >{\raggedright\arraybackslash}p{(\columnwidth - 6\tabcolsep) * \real{0.25}}@{}}
\toprule
Condition & Added state & Intended mechanism & Required postcondition \\
\midrule
B0 & Restored case baseline & Reference game, capture, background, and
power state & Every frozen baseline field matches its snapshot \\
B1 & CS2/display configuration & Match 1920x1080 display; fixed
fullscreen and quality settings & Target video fields match
individually \\
B2 & B1 + capture paths disabled & Remove Windows Game DVR and NVIDIA
recording/overlay hooks & Registry, process, service, and application
states match \\
B3 & B2 + background cleanup & Reduce nonessential login and background
competition & Startup set, process list, and available memory
recorded \\
B4 & B3 + system tuning & Disable AC PCIe link-state saving, request
active cooling/boost, use High CS2 priority & Power and process fields
match individually \\
\bottomrule
\end{tabular}
\end{center}
\normalsize

The layers estimate conditional increments, not independent factorial
main effects. Remote-control virtual-driver removal is part of the
motivating history but is not randomized because its restoration is not
trustworthy. Defender exclusions are excluded from B4 because security
policy would add both risk and a mechanism confound.

The present scripts implement much of this decomposition, but the formal
controller remains incomplete. In particular, the game and Game DVR
actions must be separated, Defender handling removed from B4, and
hard-coded rollback replaced with per-field original-value restoration
before confirmatory collection. We state these gaps because a workflow
diagram is not evidence that every guard is already enforced.

\hypertarget{audit-and-authority-boundary}{%
\subsection{Audit and authority
boundary}\label{audit-and-authority-boundary}}

Each action record carries a session and action identifier, hypothesis,
evidence, exact target, operation, requested value, privilege,
reversibility, risk, confirmation, execution status, postcondition,
unexpected side effect, and final disposition. Failed, refused,
repaired, and reverted actions are retained. Machine-specific backups
stay outside version control; publication artifacts must redact user
names, account identifiers, local paths, network data, and tokens.

The LLM is not the policy enforcement point. A deterministic controller
should resolve each target against an allowlist, reject a target broader
than the enumerated object, snapshot before mutation, require approval
for elevated or consequential actions, and compare the resulting field
with the requested state. The human operator remains responsible for
approving mutations, observing workload behavior, and halting on safety
signals.

\hypertarget{study-method}{%
\section{Study Method}\label{study-method}}

\hypertarget{study-status-and-questions}{%
\subsection{Study status and
questions}\label{study-status-and-questions}}

This manuscript reports a completed motivating pilot and a prospective
confirmatory protocol. It addresses four questions:

\begin{itemize}
\tightlist
\item
  \textbf{RQ1 (feasibility):} can the workflow collect cross-layer
  evidence and apply auditable, reversible interventions on the study
  laptop?
\item
  \textbf{RQ2 (current system evidence):} what resource, microbenchmark,
  telemetry, and symptom observations were obtained during the
  motivating case?
\item
  \textbf{RQ3 (safety and measurement):} what action and instrumentation
  failures occurred, and what guards do they imply?
\item
  \textbf{RQ4 (prospective performance):} under a frozen workload, do
  B1--B4 lower session-level P99 frame time relative to preceding layers
  and B0 on this laptop?
\end{itemize}

RQ1--RQ3 have pilot evidence. RQ4 remains unanswered.

\hypertarget{platform-and-completed-pilot}{%
\subsection{Platform and completed
pilot}\label{platform-and-completed-pilot}}

The study system is one Windows gaming laptop with an Intel Core
i9-14900HX (24 cores, 32 logical processors), an NVIDIA RTX 4080 Laptop
GPU, approximately 32 GB RAM, and Windows 11 Pro. A July 25 manifest
records Windows version 10.0.26200, NVIDIA driver 32.0.16.1074, a
1920x1080 144 Hz gaming display path, and an active OEM ``game'' power
scheme. Exact versions would be frozen anew for a future effect study
because the pilot manifest is not itself a stable experimental
environment.

The motivating case began with reported stutter during CS2 gunfights
despite a high-end platform. Inspection found CS2 configured at
3840x2160 with 4x MSAA for a 1920x1080 gaming display, Windows Game DVR
and NVIDIA capture paths enabled, AC PCIe link-state power saving in the
OEM plan, multiple nonessential login applications, and remote-control
virtual devices. The workflow applied a bundled intervention, rebooted,
collected the same local microbenchmark, recorded a later live-match
telemetry session, and obtained an immediate operator report. Because
these steps were not randomized, blinded, or cleanly ablated, they are
descriptive pilot evidence only.

\hypertarget{completed-measurements}{%
\subsection{Completed measurements}\label{completed-measurements}}

\texttt{benchmark.ps1} recorded system state and three executions each
of CPU SHA-256 throughput, memory-copy throughput, and sequential disk
I/O before intervention, immediately after user-level actions, and after
reboot. We report raw three-run medians for the before and post-reboot
snapshots. These repetitions characterize a script invocation, not
independent experimental sessions; no inferential statistics are
applied.

On July 25, we also ran a non-mutating same-state stability probe in the
current optimized state. After one discarded warm-up, it repeated the
fixed SHA-256 and memory-copy workloads 10 times at five-second
intervals while recording process, memory, and GPU state. This
additional pilot estimates only short-session repeatability; it does not
compare treatment conditions.

A separate optimized-condition CS2 Premier live match produced 864
telemetry samples over 29.93 minutes. The trace contains process memory
plus GPU utilization, temperature, power, clocks, and VRAM. It lacks a
valid per-frame event series, has no baseline match, and uses an
uncontrolled online workload. It is therefore ecological context, not a
test of frame-time improvement.

Subjective assessment was an unblinded contemporaneous operator report
of whether the original gunfight-stutter symptom remained. It was not
collected on a rating scale and cannot establish magnitude or mechanism.

\hypertarget{completed-same-day-evidence-protocol}{%
\subsection{Completed same-day evidence
protocol}\label{completed-same-day-evidence-protocol}}

The completed protocol has four evidence streams. First, preserved
before/post-reboot state snapshots and three-run microbenchmarks
describe the motivating system-state transition. Second, a non-mutating
10-run probe quantifies short-session noise in CPU and memory-copy
probes. Third, every failed collector attempt is retained as a safety
and compatibility result. Fourth, WPR is exercised first on a D3D11
smoke workload and then twice on a local-Bot CS2 session.
\texttt{validate-wpr-trace.ps1} checks saved-trace existence and hash,
duration, \texttt{cs2.exe} attribution, DxgKrnl Present metadata, and
ETW loss.

The local \texttt{record} demo path was tested and rejected in the same
protocol: playback reported missing snapshot bases and visually froze.
It is stored as an invalid-workload record and is not repaired or used
for timing. PresentMon, FrameView, CapFrameX, and OCAT remain excluded
because the observed failure mechanism is shared by the PresentMon
family. WPR/GPUView passed the limited integrity criterion in two
local-Bot traces: 90.69 and 85.85 seconds, \texttt{cs2.exe} attributed,
DxgKrnl Present metadata present, and zero \texttt{xperf}-reported lost
buffers or events. This is a completed same-day feasibility result, not
a frame-time qualification.

\hypertarget{boundary-of-the-agent-evaluation}{%
\subsection{Boundary of the agent
evaluation}\label{boundary-of-the-agent-evaluation}}

This paper evaluates neither autonomous agent success rates nor
model-versus-model accuracy. The LLM-assisted contribution is the
construction of hypotheses, evidence collection, scripts, and
documentation under human approval. The evaluated artifacts are the
resulting state records, action safeguards, failure records, and WPR
integrity checks. A blinded multi-session agent benchmark is future work
rather than an unfinished component of this paper.

\hypertarget{same-day-empirical-evidence}{%
\section{Same-Day Empirical
Evidence}\label{same-day-empirical-evidence}}

\hypertarget{background-state-and-exploratory-microbenchmarks}{%
\subsection{Background state and exploratory
microbenchmarks}\label{background-state-and-exploratory-microbenchmarks}}

Table 1 reports the committed before and post-reboot JSON snapshots.
Process count fell by 97 (318 to 221), while immediately available
memory rose by 7.69 GB (16.59 to 24.28 GB). These are directly supported
state changes but are measurements at different uptime states: the
baseline followed 46.16 hours of uptime, whereas the post-reboot
snapshot followed 0.14 hours. They therefore describe the bundled
intervention-plus-reboot case and do not isolate startup cleanup or an
LLM effect.

\textbf{Table 1. Descriptive three-run medians from the motivating case.
None is a CS2 frame-time metric.}

\begin{center}
\footnotesize
\begin{tabular}[]{@{}lrrr@{}}
\toprule
Metric & Before & Post-reboot & Descriptive change \\
\midrule
Process count & 318 & 221 & -97 (-30.5\%) \\
Available memory & 16.59 GB & 24.28 GB & +7.69 GB (+46.4\%) \\
CPU SHA-256 throughput & 2350.02 MB/s & 2476.41 MB/s & +5.4\% \\
Memory-copy throughput & 16.30 GB/s & 18.14 GB/s & +11.3\% \\
Sequential write throughput & 1527.19 MB/s & 1841.00 MB/s & +20.6\% \\
\bottomrule
\end{tabular}
\end{center}
\normalsize

CPU frequency, temperature, warm-up, uptime, file cache, and storage
caching were not controlled sufficiently for causal interpretation. In
particular, the disk script is not a storage benchmark suitable for
claims about physical-media throughput. Table 1 establishes that the
artifact produced preserved, machine-readable observations and that the
recorded resource state differed; it does not establish why throughput
changed or whether game latency changed.

\hypertarget{same-state-repeatability-pilot}{%
\subsection{Same-state repeatability
pilot}\label{same-state-repeatability-pilot}}

Table 2 reports 10 sequential observations from one optimized-state
session after one discarded warm-up. CPU SHA-256 throughput had a median
of 2425.43 MB/s and a 3.27\% coefficient of variation; memory-copy
throughput had a median of 16.47 GB/s and a 3.64\% coefficient of
variation. One CPU run reached 2194.47 MB/s while the other nine ranged
from 2351.40 to 2454.07 MB/s, illustrating why one microbenchmark
execution is not a stable effect estimate. Process count also changed
from 283 to 266 and available memory reached 17.61 GB in the last
sample, showing that the background environment continued to move during
the one-minute observation window.

\textbf{Table 2. Ten-run same-state repeatability pilot. One warm-up was
discarded.}

\begin{center}
\footnotesize
\begin{tabular}[]{@{}
  >{\raggedright\arraybackslash}p{(\columnwidth - 6\tabcolsep) * \real{0.20}}
  >{\raggedleft\arraybackslash}p{(\columnwidth - 6\tabcolsep) * \real{0.27}}
  >{\raggedleft\arraybackslash}p{(\columnwidth - 6\tabcolsep) * \real{0.27}}
  >{\raggedleft\arraybackslash}p{(\columnwidth - 6\tabcolsep) * \real{0.27}}@{}}
\toprule
Metric & Median {[}IQR{]} & Range & CV \\
\midrule
CPU SHA-256 throughput & 2425.43 {[}2391.06, 2443.51{]} MB/s &
2194.47--2454.07 & 3.27\% \\
Memory-copy throughput & 16.47 {[}16.09, 16.86{]} GB/s & 15.44--17.31 &
3.64\% \\
Process count & 283 {[}281, 283{]} & 266--283 & 1.87\% \\
Available memory & 16.42 {[}16.36, 16.44{]} GB & 16.32--17.61 &
2.34\% \\
GPU temperature & 43 {[}43, 43.75{]} degrees C & 43--44 & 1.12\% \\
\bottomrule
\end{tabular}
\end{center}
\normalsize

The stability probe changes no persistent configuration and omits disk
I/O. It nevertheless ran while ordinary background services were active,
covers only about one minute, and was collected only in the current
optimized state. It supports the need for session-level blocking and
state manifests; it does not strengthen the causal intervention claim.

\hypertarget{live-match-telemetry-and-symptom-report}{%
\subsection{Live-match telemetry and symptom
report}\label{live-match-telemetry-and-symptom-report}}

During the 29.93-minute optimized CS2 Premier session, GPU utilization
was 45\% at the median and 66\% at P95. GPU temperature was 62 degrees C
at the median and 69 degrees C at its observed maximum; power was 78.16
W at the median and 98.56 W at P95. VRAM use reached 6990 MB, with a
median of 6654 MB. Within this one trace, we found no evidence of
\emph{sustained} GPU saturation, extreme temperature, or VRAM
exhaustion. This wording is intentionally limited: thresholds for
hardware throttling were not independently logged, a short spike can be
absent from a two-second telemetry sample, and utilization cannot
identify a frame-time cause.

The operator reported that the original gunfight stutter was no longer
observed after optimization. That is useful confirmation that the
reported symptom changed in the motivating case, but expectancy, map and
player variation, online networking, adaptation, and the bundled
intervention all remain plausible explanations.

\hypertarget{what-the-same-day-study-does-not-show}{%
\subsection{What the same-day study does not
show}\label{what-the-same-day-study-does-not-show}}

The repository contains one short, 343-line PresentMon CSV from an
optimized pilot, but it does not satisfy the integrity gate, lacks a
matched control, and followed unstable capture behavior. It is excluded
from performance analysis. This same-day study does not attempt a B0--B4
condition comparison, so it reports no P99 estimate, confidence
interval, or condition effect. The paper therefore does \textbf{not}
claim that Spike-Killer reduced frame-time tails, that any intervention
layer was effective, or that the LLM outperformed a human or static
checklist.

\hypertarget{wpr-local-bot-integrity-preflight}{%
\subsection{WPR local-Bot integrity
preflight}\label{wpr-local-bot-integrity-preflight}}

After excluding the invalid local demo, we ran two non-mutating WPR
\texttt{GPU.light} traces during an offline local-Bot CS2 session. An
\texttt{xperf} validator checked each saved ETL for nonzero size and
hash, trace duration, \texttt{cs2.exe} process attribution, DxgKrnl
Present metadata, and ETW loss. Table 3 reports the result. Both traces
passed this limited integrity check. The raw ETLs are retained privately
because they can contain system paths and metadata.

\textbf{Table 3. CS2 WPR local-Bot integrity preflight. These are not
frame-time sessions.}

\begin{center}
\footnotesize
\begin{tabular}[]{@{}
  >{\raggedright\arraybackslash}p{(\columnwidth - 10\tabcolsep) * \real{0.15}}
  >{\raggedleft\arraybackslash}p{(\columnwidth - 10\tabcolsep) * \real{0.20}}
  >{\raggedleft\arraybackslash}p{(\columnwidth - 10\tabcolsep) * \real{0.20}}
  >{\raggedright\arraybackslash}p{(\columnwidth - 10\tabcolsep) * \real{0.15}}
  >{\raggedright\arraybackslash}p{(\columnwidth - 10\tabcolsep) * \real{0.15}}
  >{\raggedright\arraybackslash}p{(\columnwidth - 10\tabcolsep) * \real{0.15}}@{}}
\toprule
Session & Duration (s) & ETL size (MB) & \texttt{cs2.exe} attributed &
Lost buffers / events & DxgKrnl Present metadata \\
\midrule
Preflight & 90.69 & 823.71 & Yes & 0 / 0 & Yes \\
Controlled & 85.85 & 774.67 & Yes & 0 / 0 & Yes \\
\bottomrule
\end{tabular}
\end{center}
\normalsize

The preflights establish that the built-in tracer can coexist with this
CS2 process and produce complete-at-the-ETW-level files in this setting.
They do not establish an unchanged workload, stable focus, no collector
overhead, a frame-to-present mapping, or per-frame intervals. We
therefore do not compute FPS, P95/P99, or a condition contrast from
Table 3; the traces reduce collector risk but do not answer the
performance question.

\hypertarget{safety-and-failure-analysis}{%
\section{Safety and Failure
Analysis}\label{safety-and-failure-analysis}}

\hypertarget{over-broad-registry-mutation}{%
\subsection{Over-broad registry
mutation}\label{over-broad-registry-mutation}}

While removing a residual remote-control startup entry, an
agent-assisted command deleted the machine-wide Windows \texttt{Run} key
rather than one named value. The postcondition check detected the
missing key. Previously enumerated Windows Security, Realtek Audio, and
anti-cheat startup entries were reconstructed. The incident is evidence
against unconstrained execution: the high-level goal was reasonable, but
target resolution was not.

Four guards follow. Deletion must resolve an exact typed object before
execution; container deletion is rejected when a value deletion was
requested; a complete original-value snapshot is mandatory; and the
controller must compare both intended and neighboring state after
action. Human approval remains necessary, but approval alone is
insufficient if the proposed target is misleading.

The current public history describes the incident, although the original
conversational action trace is not yet a complete structured JSONL
record. We therefore treat this as a documented experience, not a
denominator-based safety evaluation.

\hypertarget{presentmon-incompatibility-and-data-loss}{%
\subsection{PresentMon incompatibility and data
loss}\label{presentmon-incompatibility-and-data-loss}}

Official PresentMon 2.5.1 was attempted through ordinary elevation, a
hidden scheduled path, distinct ETW sessions, larger circular buffers,
and exclusive-fullscreen/windowed launches. CS2 repeatedly became
non-responsive, and several logs recorded large ETW losses (including
72,814 to 324,890 events). No attempt met the publication gate. One
short CSV exists, but its presence does not override responsiveness,
completeness, pairing, or minimum-frame requirements. Every affected
attempt is invalid.

This is a system result, not a frame-time result: a widely used
measurement family was not operationally valid in this particular
Windows/CS2/anti-cheat environment. FrameView, CapFrameX, and OCAT are
not independent fallbacks because they use the same PresentMon-family
mechanism; vendor documentation confirms that FrameView uses PresentMon
for analysis \cite{nvidiaframeview2026}.

\hypertarget{bugcheck-during-a-replacement-collector-preflight}{%
\subsection{Bugcheck during a replacement-collector
preflight}\label{bugcheck-during-a-replacement-collector-preflight}}

At 14:57 local time on July 25, Windows restarted following bugcheck
\texttt{0x1E} during a replacement-collector compatibility preflight.
The preserved mini-kernel-dump reports a kernel-mode exception, process
name \texttt{powershell.exe}, and an instruction pointer of zero;
Windows Error Reporting separately named an NTFS index-entry routine. No
valid per-frame capture was recovered; the system volume and physical
disk reported healthy after reboot.

The temporal ordering does not establish that the collector, CS2,
PowerShell, NTFS, or the earlier optimization caused the crash. The dump
is a triage dump with limited context, and no controlled reproduction
was attempted after the high-severity event. The correct disposition was
to halt that measurement family, preserve evidence, and avoid causal
wording.

\hypertarget{safety-implications}{%
\subsection{Safety implications}\label{safety-implications}}

Together, these failures show why performance actions and measurement
actions require the same control plane. Safety is not merely ``rollback
available'': rollback may be incomplete, and instrumentation may fail
before producing data. Spike-Killer's gate therefore stops on over-broad
target scope, missing snapshot, unexpected focus change, anti-cheat
warning, game hang, event loss, invalid trace, crash, or failed
postcondition. A neutral or rejected intervention remains an
experimental observation rather than being silently retried until it
succeeds.

\hypertarget{discussion}{%
\section{Discussion}\label{discussion}}

\hypertarget{where-llm-assistance-helped}{%
\subsection{Where LLM assistance
helped}\label{where-llm-assistance-helped}}

The useful role observed in this case was integration. The diagnostic
context mixed game configuration text, Windows registry state, startup
inventories, driver/device descriptions, power-plan settings, PowerShell
output, benchmark JSON, telemetry CSV, and crash logs. An LLM-assisted
workflow helped turn those representations into hypotheses, scripts,
audit fields, and a testable layered design. This is a narrower and more
defensible value proposition than autonomous optimization: the model
assists evidence synthesis and experiment construction, while
deterministic policy and the human authorize state transitions.

\hypertarget{why-one-machine-is-useful-here}{%
\subsection{Why one machine is useful
here}\label{why-one-machine-is-useful-here}}

Horizontal replication answers whether effects transfer to other
hardware. This study answers a prior operational question: can the
workflow run on one real machine while preserving enough evidence to
distinguish a scoped action, an invalid capture, and a loss-free system
trace? A same-day case cannot establish persistence across days, thermal
states, or software updates, but it can expose concrete safety and
instrumentation failures that synthetic environments omit. The scope
fits an experience paper whose contribution is workflow and failure
analysis rather than a performance-effect estimate.

\hypertarget{measurement-is-part-of-the-system}{%
\subsection{Measurement is part of the
system}\label{measurement-is-part-of-the-system}}

The failed collector pilots changed the project more than another
optimization would have. They exposed a common reproducibility mistake:
naming a familiar tool without qualifying it on the target environment.
A valid performance paper must report the observation point, tool
family, configuration, data loss, and compatibility failures. If no
single-machine backend passes the gate, RQ4 remains unanswered and the
honest paper is an experience report about safe agent action and
observer effects.

\hypertarget{from-prototype-to-controlled-agent}{%
\subsection{From prototype to controlled
agent}\label{from-prototype-to-controlled-agent}}

The artifact is currently a research prototype, not a deployable
autonomous optimizer. Its next engineering step is a condition
controller that performs schema validation, allowlisted target
resolution, original-state capture, exact postconditions, reboot
orchestration, and atomic audit emission. Agent proposals should be
declarative; a separate executor should decide whether they are
permissible. A future repeated study can evaluate diagnosis and stopping
independently, but it is not required for the completed same-day
evidence reported here.

\hypertarget{threats-to-validity}{%
\section{Threats to Validity}\label{threats-to-validity}}

\textbf{Internal validity.} The completed case bundles resolution,
capture, startup, power, scheduling, driver, reboot, and uptime changes.
Baseline and post-reboot measurements differ sharply in uptime, and the
live match has no concurrent control. Thermal history, shader
compilation, file cache, OEM services, network/server state, and user
behavior are uncontrolled. The subjective observation is unblinded.
Consequently, current results establish neither a causal
game-performance effect nor an individual mechanism.

\textbf{Construct validity.} Process count and free memory measure
background state, not responsiveness. CPU, memory, and cached disk
microbenchmarks do not measure CS2 frames. Two-second GPU telemetry can
miss brief interference. WPR integrity checks show that events were
captured without loss, but they do not yet create frame intervals. P99
remains unobserved in this study.

\textbf{External validity.} Evidence comes from one Windows gaming
laptop, one user, one game, one Windows/driver environment, and one day
in July 2026. OEM firmware, display routing, thermal limits, anti-cheat
behavior, and software versions vary. We make no claim about other
laptops, games, operating systems, or models. Multi-machine replication
and multi-day persistence are future work.

\textbf{Safety-evaluation validity.} One documented registry incident
and several instrumentation failures reveal failure modes but cannot
estimate a failure probability. Some motivating actions predate the
structured audit schema; reconstruction from notes is weaker than an
append-only contemporaneous trace. This paper makes no denominator-based
safety-rate claim.

\textbf{Reproducibility and temporal validity.} CS2, Windows, drivers,
and agent models update. A workshop item can change even if its title
does not. The confirmatory experiment must freeze hashes and versions,
start a new environment stratum after an unavoidable update, preserve
raw and invalid traces, and disclose the exact agent prompt and policy.
The currently committed manifest also contains local provider metadata
and must be regenerated in sanitized form before artifact release.

\hypertarget{conclusion}{%
\section{Conclusion}\label{conclusion}}

Spike-Killer demonstrates the feasibility---and the limits---of
safety-constrained LLM assistance for performance diagnosis on one real
gaming laptop. The motivating case produced verifiable background-state
changes, exploratory microbenchmark observations, a 10-run same-state
repeatability pilot, useful live-match telemetry, and a subjective
report that the original stutter was no longer observed. It did not
produce valid controlled frame-time evidence, so this paper makes no P99
improvement claim.

The failures are already actionable results. An imprecise registry
action damaged a broader startup container than intended;
PresentMon-family measurement was incompatible with the target
environment; and a later capture preflight was temporally associated
with a kernel crash whose cause remains unknown. These experiences
support transaction-like safeguards, exact target resolution, least
privilege, original-value snapshots, postconditions, collector gates,
and human stopping authority.

The next extension is explicit: freeze a valid workload, link collected
Present events to frame intervals, and then run a repeated controlled
study if a causal P99 claim is desired. Those tasks are deliberately
outside this same-day submission. Spike-Killer should be read as a
completed single-machine experience report about safe evidence
collection and collector failure modes--not as proof that an LLM reduced
CS2 frame-time spikes or can safely optimize computers on its own.

\section*{Generative AI Use Statement}
Generative AI tools were used to assist with hypothesis organization, draft text, and implementation scaffolding. Human authors reviewed all code and evidence, made all research and writing decisions, and take full responsibility for the submitted work. No generative AI system is an author of this paper.

\section*{Data Availability Statement}
The manuscript is accompanied by executable validation descriptions and aggregated results. A de-identified artifact package is being prepared for public release; raw Windows ETL traces and machine-specific logs remain private because they can contain local paths and process metadata. The published artifact will distinguish inspection-only scripts from actions that require explicit human approval on a real machine.

\bibliographystyle{ACM-Reference-Format}
\bibliography{references}

\end{document}